\documentstyle[12pt]{article}
\newcommand {\omutau} {$\nu_\mu \rightarrow \nu_\tau$}
\newcommand {\xlepdec} {$\tau^- \rightarrow l^-\nu\bar{\nu}$}
\newcommand {\xedec} {$\tau^- \rightarrow e^-\nu\bar{\nu}$}
\newcommand {\xmudec} {$\tau^- \rightarrow \mu^-\nu\bar{\nu}$}
\newcommand {\xpidec} {$\tau^- \rightarrow \pi^-\nu$}
\newcommand {\xrhodec} {$\tau^- \rightarrow \rho^-\nu$}
\def\xa1dec{$\tau^- \rightarrow a_1^-\nu$}
\def\yrho_2pi{$\rho^- \rightarrow \pi^-\pi^0$}
\def\ya1_3pi{$a_1^- \rightarrow \pi^-\pi^+\pi^-$}
\def\ypi_gg{$\pi^0 \rightarrow \gamma\gamma$}
\begin{document}
\title{Detecting the (Quasi-)Two-Body Decays of $\tau$ Leptons in
Short-Baseline Neutrino Oscillation Experiments}
\author{
\\A. Asratyan$^{a,}$\thanks{Corresponding author. Tel.: (095)-237 0079. 
E-mail: asratyan@vitep5.itep.ru.},
M. Balatz$^a$,
D. Boehnlein$^b$,
S. Childres$^b$,\\
G. Davidenko$^a$, 
A. Dolgolenko$^a$,
G. Dzyubenko$^a$,\\
V. Kaftanov$^a$,
M. Kubantsev$^{a,}$\thanks{Now at Kansas State University, Manhattan, 
KS 66506, USA},
N. W. Reay$^c$,
J. Musser$^d$,\\
C. Rosenfeld$^e$,
N. R. Stanton$^c$,
R. Thun$^f$,
G. S. Tzanakos$^g$,\\
V. Verebryusov$^a$,
and V. Vishnyakov$^a$\\ 
\normalsize
{\it (a) Institute of Theoretical and Experimental Physics, 117259 Moscow, 
Russia} \\
\normalsize
{\it (b) Fermi National Accelerator Laboratory, Batavia, IL 60510, USA} \\
\normalsize
{\it (c) Kansas State University, Manhattan, KS 66506, USA} \\
\normalsize
{\it (d) Indiana University, Bloomington, IN 47405, USA} \\
\normalsize
{\it (e) University of South Carolina, Columbia, SC 29208, USA} \\
\normalsize
{\it (f) University of Michigan, Ann Arbor, MI 48109, USA} \\
\normalsize
{\it (g) University of Athens, 15771 Athens, Greece}
}                           % end of Author
%\date {\today}
\maketitle
\begin{abstract}
Novel detector schemes are proposed for the short-baseline neutrino 
experiments of next generation, aimed at exploring the large-$\Delta m^2$ 
domain of \omutau\ oscillations in the appearance mode. These schemes
emphasize good spectrometry for charged particles and for electromagnetic 
showers and efficient reconstruction of \ypi_gg\ decays. The basic elements
are a sequence of relatively thin emulsion targets, immersed in magnetic
field and interspersed with electronic trackers, and a fine-grained 
electromagnetic calorimeter built of lead glass. These elements 
act as an integral whole in reconstructing the electromagnetic showers.
This conceptual scheme shows good performance in identifying the  $\tau$
(quasi-)two-body decays by their characteristic kinematics and in selecting
the electronic decays of the $\tau$.
\end{abstract}
\newpage
\section{Introduction}
     Whether or not the leptonic number is strictly conserved is still 
an open problem of particle physics. The leptonic number may be 
violated, if the current neutrinos $\nu_\alpha$ ($\alpha = e, ~\mu, ~\tau$) 
are not exactly massless, but rather represent linear combinations of the
mass eigenstates  $\nu_i$ with masses  $m_i$ ($i$ = 1, 2, 3):  
$\nu_\alpha = U_{\alpha i} \nu_i$, where $U_{\alpha i}$ is a unitary mixing 
matrix that is analogous to the Cabibbo-Kobayashi-Maskawa matrix for the quark
sector. Then, as different mass components no longer propagate coherently, 
one current neutrino may oscillate in flight to another current 
neutrino \cite{prophets}: $\nu_\alpha \rightarrow \nu_\beta$, 
$\alpha \neq \beta$. Over a distance $L$ from the point of emission, the
transition probability is
\begin{displaymath}
P(\alpha \rightarrow \beta) = \delta_{\alpha\beta} - 
4 \sum_{i > j} U_{\alpha i} U_{\beta i} U_{\alpha j} U_{\beta j}
\sin^2(\pi L / \lambda_{ij}) ,
\end{displaymath}
where the oscillation length, $\lambda_{ij}$, is proportional to the
neutrino energy  $E_\nu$  and is inversely proportional to the mass 
difference  $\Delta m^2_{ij} = m_i^2 - m_j^2$ :
\begin{displaymath}
%  \lambda_{ij} = 4 \pi E_\nu \bar{h} c / \mid \Delta m^2_{ij} \mid .
\lambda_{ij} = 4 \pi E_\nu \hbar c / \mid \Delta m^2_{ij} \mid .
\end{displaymath}
Note that in the considered case of just three Dirac neutrinos, we have
only two independent mass differences  $\Delta m^2_{ij}$. Given the values of
$E_\nu$ and  $\Delta m^2$, the transition probability reaches its first 
maximum at a distance $L = 1.24$ km $E_\nu$ (GeV) / $\Delta m^2$ (eV$^2$)
from the point of emission.

     Searches for neutrino oscillations are actively pursued in experiments
with solar, atmospheric, reactor, and accelerator neutrinos. These are
characterized by very different $L/E_\nu$ ratios and, therefore, are
sensitive to very different values of $\Delta m^2$. A large deficit of muon 
neutrinos, as observed in atmospheric showers \cite{atmospheric}, suggests
that they transform to another flavor with $\Delta m^2$ on the order of 
10$^{-2}$--10$^{-3}$ eV$^2$  and with an effective mixing, defined as
$\sin^2 (2\theta) = 4 U_{\alpha i} U_{\beta i} U_{\alpha j} U_{\beta j}$,
between 0.5 and 1.0. On the other hand, an accelerator experiment using 
neutrinos from $\pi^+$ and $\mu^+$ decays at rest \cite{lsnd} has presented 
evidence for the  $\bar{\nu}_\mu \rightarrow \bar{\nu}_e$  transition with  
a much larger mass difference of $\Delta m^2 \sim 1$ eV$^2$ and with a much
smaller effective mixing of  $\sin^2 (2\theta) \sim 6 \times 10^{-3}$.

     The former pattern of flavor-changing transitions between neutrinos, 
that is characterized by a small  $\Delta m^2$  and a large mixing, will
be thoroughly investigated in long-baseline accelerator experiments like
MINOS at Fermilab \cite{minos}, in which the muon neutrinos will travel
a distance of 731 km before hitting the detector. The latter pattern of
oscillations, that involves  $\Delta m^2 \sim 1$ eV$^2$  and a small mixing,
must be further studied in either the  $\nu_\mu \rightarrow \nu_e$  and
$\nu_\mu \rightarrow \nu_\tau$  channels in short-baseline accelerator
experiments that may accumulate sufficient statistics of neutrino
interactions.

     Despite some appealing attempts to describe all available data as
a whole \cite{perkins, pakvasa}, we believe that the experimental situation 
is still too volatile to allow a coherent and reliable scheme of transitions
among the three neutrino flavors. Still, there are sound qualitative reasons 
to believe that the transition driven by the bigger $\Delta m^2_{ij}$ must 
primarily manifest itself in the  $\nu_\mu \rightarrow \nu_\tau$  channel.
Qualitatively, one may expect that the mass hierarchy of neutrinos follows 
that of corresponding charged leptons, so that $\nu_\tau$ should be the most 
massive. Indeed, the GUT see-saw mechanism ~\cite{seesaw} predicts that  
$m_{\nu_e} / m_{\nu_\mu} / m_{\nu_\tau} \sim m_u^2 / m_c^2 / m_t^2$. 
On the other hand, in analogy with the known pattern of quark mixings, one may 
naively expect that mixings between the neighboring neutrino flavors 
({\it i.e.}, $\nu_e$--$\nu_\mu$  and  $\nu_\mu$--$\nu_\tau$) should be the 
strongest, while that for $\nu_e$--$\nu_\tau$  may be substantially weaker. 

     Therefore, if the bigger  $\Delta m^2_{ij}$  is on an order of 1 eV$^2$ 
as suggested by the data of LSND \cite{lsnd}, one should look for the 
corresponding $\nu_\mu \rightarrow \nu_\tau$  transition in the short-baseline 
($L \sim 1$ km) and medium-baseline ($L \sim 10-20$ km) experiments
at accelerators. Quite importantly, the accelerators like SPS at CERN and
Main Injector at Fermilab offer intense and almost pure beams of muon 
neutrinos with mean energies well above the threshold for $\tau$ production, 
as required by experiments with $\nu_\tau$ appearance.

     The flux of $\tau$ neutrinos generated via the flavor-changing 
transition $\nu_\mu \rightarrow \nu_\tau$  can only be inferred from the 
number of $\tau$ leptons produced in the charged-current reaction  
$\nu_\tau N \rightarrow \tau^- X$. The created $\tau$ typically travels a 
distance of one millimeter in space (lifetime $3\times10^{-13}$ s) and 
then decays either leptonically with emission of two neutrinos  
($\tau^- \rightarrow \mu^- \bar{\nu} \nu$, $e^- \bar{\nu} \nu$)  or 
hadronically with emission of a single neutrino 
($\tau^- \rightarrow \pi^- \nu$, $\rho^- \nu$, $a_1^- \nu$, {\it etc.}). An 
efficient search for the  $\nu_\mu \rightarrow \nu_\tau$
transition requires that the events of $\tau$ production and decay be selected 
and analyzed on an event-by-event basis. Therefore, the production and 
decay vertices must be reliably resolved in the detector (whereby the short 
track of the $\tau$ is reconstructed in space) and the decay particles must be 
identified. Of the available detector choices, only nuclear emulsion fully 
meets the above requirements for a massive target. 

     The emulsion technique was pioneered by the Fermilab experiment E531 
which, in its search for the  $\nu_\mu \rightarrow \nu_\tau$  
transition \cite{e531limit} in the appearance mode, found no candidate events 
of $\tau$ production and decay. 
For  $\Delta m^2 > 50$ eV$^2$, the mixing parameter  $\sin^2 (2\theta)$ was 
restricted to be less than $5\times 10^{-3}$. At CERN SPS, the emulsion
experiment CHORUS \cite{chorus} completed operation in 1997, but data
processing is still in progress \cite{progress}. Provided that no candidate 
events are finally seen, they will tighten the E531 upper limit on  
$\sin^2 (2\theta)$ by another order of magnitude.
\section{Short-Baseline Experiments of Next Generation}
     Two short-baseline emulsion experiments of the next generation have been 
designed to achieve an order-of-magnitude increase in sensitivity over CHORUS 
\cite{chorus} and the electronic experiment NOMAD \cite{nomad}. At Fermilab,
E803/COSMOS \cite{cosmos1, cosmos2} was optimized for the intense neutrino 
beam of Main Injector -- a 120-GeV proton accelerator to be commissioned
in 1999. At CERN, TOSCA \cite{tosca} will use the $\nu_\mu$ beam generated
by the 350-GeV proton beam of the SPS accelerator. Like E531 and CHORUS, both
are designed as hybrid detectors combining an emulsion target with a 
downstream electronic spectrometer. The candidates for $\tau$ decays are 
directly observed in the emulsion target, while the spectrometer is used to 
select the events to be scanned in emulsion and to measure the momenta of 
secondary particles for kinematic analysis of $\tau$ candidates. 

     The baseline schemes of COSMOS and TOSCA are detailed in the respective
proposals \cite{cosmos2} and \cite{tosca}, and general characteristics of the 
two experiments are compared in Table \ref{compare}. Either experiment will 
collect ten times more CC interactions of muon neutrinos than CHORUS, while 
rejecting the backgrounds to the $\tau$ signal much more efficiently. The 
bigger target mass and higher beam energy of TOSCA are seen to be largely 
compensated by the higher repetition rate of Main Injector that will deliver 
to target some eight times more protons per year than CERN SPS. As a result, 
the two experiments may boast comparable sensitivities to the  
$\nu_\mu \rightarrow \nu_\tau$ transition in the null limit, that is, 
assuming that no signals are finally detected.

     That the experiment be capable of convincingly interpreting and 
demonstrating even a relatively small $\tau$ signal, should one show up,
is perhaps much more important than just restricting the parameter space
for the null hypothesis. For this, an appreciable fraction of the
signal must be unambiguously reconstructed as $\tau$ decays rather than
the decays of anticharm produced by the antineutrino component of the beam.
Such distinctive signatures can only be provided by the (quasi-)two-body
semileptonic decays of the $\tau$ :
\begin{displaymath}
\tau^- \rightarrow \pi^- \nu ~(BR = 11.3\%) , 
\end{displaymath}
\begin{displaymath}
\tau^- \rightarrow \rho^- \nu, ~\rho^- \rightarrow \pi^- \pi^0 ~(BR = 25.2\%),
\end{displaymath}
\begin{displaymath}
\tau^- \rightarrow a_1^-\nu, ~a_1^-\rightarrow\pi^-\pi^+\pi^- ~(BR = 9.4\%).
\end{displaymath}
In a (quasi-)two-body decay  $\tau^- \rightarrow h^-\nu$, "transverse mass" 
is defined as $M_T = \sqrt {m_h^2 + p_T^2} + p_T$, where $m_h$ and $p_T$ 
are the $h^-$ mass and transverse momentum with respect to $\tau$ direction.  
The two-body kinematics dictate that the $M_T$ distribution should reveal a
very distinctive peak just below  $M_T^{max} = m_\tau$. This "Jacobian cusp",
that may provide a characteristic signature of the $\tau$, rapidly degrades 
with increasing $\Delta p / p$ for the decay products (see \cite{cosmos1, 
cosmos2} and the figures below). Needless to say, analyzing the \xrhodec\
decays also requires good reconstruction of \ypi_gg\ decays in the detector. 
The $M_T$ technique for identifying massive parents by two-body decays in 
emulsion was proven by observing the relatively rare decays
$D_s^+(1968) \rightarrow \mu^+ \nu$  against a heavy background from other
decays of charm \cite{e653}. 

     In its baseline form that emphasizes good spectrometry for charged
particles ($\Delta p / p \sim 0.03$ disregarding the error from multiple
scattering in emulsion), COSMOS will be able to observe a distinct
Jacobian cusp in the $M_T$ distribution for the pionic decay  \xpidec. The
decay \xrhodec\ will be analyzed less efficiently because of difficulties
in reconstructing the \ypi_gg\  decays in thick emulsion (three radiation
lengths in the baseline design). Thick emulsion will also complicate the
analysis of electronic decays
\begin{displaymath}
\tau^- \rightarrow e^- \bar{\nu} \nu ~(BR = 17.8\%)
\end{displaymath}
and subsequent comparison with the muonic decays
\begin{displaymath}
\tau^- \rightarrow \mu^- \bar{\nu} \nu ~(BR = 17.4\%).
\end{displaymath}

     In TOSCA, the analysis of (quasi-)two-particle decays of the $\tau$ is 
severely hampered by inferior spectrometry for charged particles 
($\Delta p / p \sim 0.1$) and by the fact that the decays \ypi_gg\ are not 
reconstructed at all. The one-prong semileptonic decays
\begin{displaymath}
\tau^- \rightarrow \pi^-(n\pi^0)\nu ~(BR \sim 50\%)
\end{displaymath}
are treated only inclusively, {\it i.e.}, without selecting the individual 
(quasi-)two-particle channels. In this sense, in its baseline form TOSCA
is a "counting experiment" that only compares the number of observed kinks 
with predicted backgrounds from anticharm decays and from pion interactions
in emulsion. Thus, despite a somewhat higher sensitivity to the
$\nu_\mu \rightarrow \nu_\tau$  transition in the null limit, 
TOSCA may have less discovery potential than COSMOS.
\section{Alternative Schemes for COSMOS and TOSCA}
     In this paper, we propose alternative conceptual schemes for COSMOS and
TOSCA, aimed at enhancing the discovery potentials of both experiments through
improving the analysis of (quasi-)two-body decays of the $\tau$. Accordingly,
the proposed schemes emphasize good spectrometry for charged particles and
electromagnetic showers and efficient reconstruction of \ypi_gg\ decays.

     Either scheme involves a sequence of relatively thin emulsion targets,
immersed in magnetic field and interspersed with electronic trackers,
and a fine-grained electromagnetic calorimeter (EMCal) 
downstream of the targets. As the total thickness of emulsion amounts
to several radiation lengths, most photons originating from the targets will 
produce electromagnetic showers. The adopted strategy is to reassemble the
parent photon from unconverted daughter gammas (detected in the EMCal) and
from conversion electrons that are momentum-analyzed in the instrumented gaps.
By additionally sampling the shower in between the targets that act as 
converters, we are able to collect a significant fraction of parent energy 
despite relatively large total thickness of emulsion. Thus, in reconstructing
electromagnetic showers the targets, trackers, and the EMCal act as an
integral whole. This strategy develops the original philosophy of COSMOS and
is very alien to the baseline design of TOSCA in which the six target modules
are virtually independent of each other \cite{tosca}.

     The strategy of assembling the parent photon from "pieces" dictates that 
the EMCal should show good performance at energies down to some 100 MeV in a 
high-multiplicity environment (10--20 showers per event). Of the existing
options, these requirements are best met by the technique of Cherenkov
lead glass. For either COSMOS and TOSCA, we assume a round-shaped EMCal with
3.5-m diameter, built of lead-glass cells with transverse granularity of
4.25 by 4.25 cm and with thickness of 15 radiation lengths. The cells are read 
out by photomultipliers. We have ample experience in constructing and 
operating very similar devices, and the response of the proposed EMCal is 
fully understood through extensive simulations and calibrations 
\cite{leadglass}. It is characterized by an energy resolution of
$\delta E$ (GeV) = $0.05 \sqrt{E} + 0.02 E$, while the position resolution
in either coordinate is close to 2 mm.

     In the proposed schemes, as in the baseline design of COSMOS, electronic 
tracking largely relies on the multisampling (jet) drift chambers that provide 
excellent pattern recognition, redundancy, and two-track resolution. Following
the original design of TOSCA, we also include two planes of silicon-microstrip
detectors, providing two independent $XY$ measurements to define a track 
segment, just downstream of each target (at 10 and 50 mm from bulk emulsion,
respectively). These are primarily aimed at facilitating the extrapolation of 
tracks into bulk emulsion \cite{scanback}, but also affect the spectrometry
of these tracks. In the simulations described in subsequent sections, we 
assume that the drift chambers and silicon-microstrip planes have spatial 
resolutions of 150 and 20 $\mu$m, respectively.

     In either scheme detailed below, we assume that the targets and trackers
are immersed in uniform magnetic field of 0.5 Tesla that is perpendicular to 
beam direction. The geometry of the magnet is not specified. (Note however
that TOSCA will use the existing dipole magnet with useful volume of
$3.5\times3.5\times7.5$ m$^3$  and field of up to 0.7 Tesla, that fully meets
our requirements. At the same time, our scheme for COSMOS will require a 
bigger magnet than that foreseen in the original proposal.) Likewise, we
do not specify the geometry of the muon identifier that is not relevant to 
our analysis.

     The proposed configuration of COSMOS is shown in Fig. \ref{detectors}.
The three emulsion targets with areas of  $180\times180$ cm$^2$  and 
thicknesses of 3 cm ($\sim$ one radiation length) are located at
$z =$ 0, 120, and 240 cm  along the beam direction. The front face of the
EMCal is at  $z = 540$ cm, or far enough from the targets for efficiently
reconstructing the \ypi_gg\ decays. Also shown are the pairs of 
silicon-microstrip planes just downstream of each target and the drift
chambers with active area of $240\times240$ cm$^2$ that instrument the
120-cm-wide gaps between the targets and the 300-cm-wide gap between the
downstream target and the EMCal. The total thickness of emulsion is the
same as in the baseline scheme (9 cm), but now we are able to sample 
electromagnetic showers every radiation length of emulsion. The total amount 
of emulsion is 1112 kg, or some 30\% more than in the baseline design.

     For TOSCA, we propose a very similar configuration that is also depicted 
in Fig. \ref{detectors}. Here, the number of targets is increased to four
(at  $z =$ 0, 150, 300, and 450 cm) and their thicknesses are increased
to 4 cm, thus bringing the total thickness of emulsion to over five radiation
lengths. Compared to COSMOS that will operate at a lower beam energy
(see Table \ref{compare}), the targets are driven farther apart to provide
a longer lever arm for momentum analysis.
The targets and the drift chambers have the same areas as in the
COSMOS configuration ($180\times180$ and $240\times240$ cm$^2$, 
respectively). That the EMCal is now placed farther downstream of the last 
target (front face at  $z = 950$ cm) is dictated by higher mean energy of 
the $\pi^0$ mesons to be reconstructed by TOSCA. Compared to the baseline 
design of TOSCA, the total amount of emulsion is less by some 17\%, while 
the total area of silicon-microstrip planes is virtually the same.
\section{Simulating $\tau$ Signatures in the Detector}
     For either scheme detailed in the previous section, we have performed
GEANT-based simulations of detector response to the process of $\tau$
production and decay through various channels. In our calculations, we
assume that the energy spectra of incident $\tau$ neutrinos are exactly
proportional to those of original $\nu_\mu$ beams from the respective
accelerators ({\it i.e.}, from Main Injector for COSMOS and from CERN SPS 
tuned to 350-GeV proton energy for TOSCA, see Table \ref{compare}). This 
corresponds to  \omutau\  oscillations driven by a large mass difference 
($\Delta m^2 > 30$ and 50 eV$^2$ for COSMOS and TOSCA, respectively). The
radial distribution of either beam at detector position is taken into
account. The computed energy spectra of $\nu_\tau$-induced CC interactions 
for either experiment, that are affected by the energy threshold for $\tau$
production, are shown in Fig. \ref{cc_spectra}.

     Note that a realistic simulation should include the experimental 
restrictions dictated by the emulsion technique:
\begin{itemize}
\item
Because of Coulomb scattering, soft and broad-angle tracks can neither be
reconstructed in emulsion nor be linked to the tracker. Therefore, a charged 
daughter of the $\tau$ can only be detected and momentum-analyzed, if its 
lab. momentum exceeds some 1 GeV and if its emission angle is within
some 400 mrad.
\item
In a one-prong decay (like \xlepdec, \xpidec, or \xrhodec), the angle
between the $\tau$ and charged-daughter directions must exceed some 10 mrad
for the "kink" on the $\tau$ track to be detected in emulsion.
\item
At low transverse momenta, the bulk of detected kinks are due to decays
of strange particles and to pion scatterings on the nuclei of emulsion 
without visible breakup of the nucleus (the so-called "white stars").
These background kinks are largely removed by the cut  $p_T > 250$ MeV  on
transverse momentum of the charged daughter with respect to parent-$\tau$ 
direction.
\end{itemize}

     A major source of background to the $\tau$ signal is anticharm 
production in those antineutrino-induced CC events in which the primary
charged lepton (either a $\mu^+$ or a $e^+$) has been misidentified in the
detector. That the $\tau$, unlike anticharm, is emitted back-to-back with
primary hadrons in the transverse plane allows to suppress the anticharm 
background by additional kinematic cuts (see \cite{cosmos1, cosmos2, tosca})
that are not discussed in this paper.
\subsection{Simulation Procedure}
     The first stage is to generate the CC collisions of $\tau$ neutrinos 
with free nucleons, $\nu_\tau N \rightarrow \tau^- X$, with subsequent 
leptonic and semileptonic (quasi)-two-particle decays of the emitted $\tau$: 
$\tau^-  \rightarrow  l^- \nu \bar{\nu}$, $\pi^- \nu$, $\rho^- \nu$,  
and $a_1^- \nu$. Both the quasielastic and deep-inelastic processes of $\tau$ 
production are simulated. For quasielastics, the hadronic tensor is expressed 
through the known formfactors for the weak $n \rightarrow p$ transition. 
(Excitation of baryonic resonances like $\Delta^+$ and $\Delta^{++}$ is not 
simulated explicitly, but is accounted for by simply scaling up by a factor 
of 2.2 the cross section for $\nu_\tau N  \rightarrow  \tau^- N'$  on an 
isospin-averaged nucleon.) For inelastic production of the $\tau$, we rely on 
the naive parton model.

     The process of $\tau$ production and decay is treated as a whole
using the narrow-width approximation: for each decay channel, the cross
section is computed from an overall Feynman graph with two weak vertices
in which $\delta (q^2 - m_\tau^2)$ is substituted for the $\tau$ propagator.
In this way, we are able to account for polarization of the $\tau$ that
strongly affects the angular distributions of secondary particles in 
the $\tau$ frame: thus, in the decay  $\tau^- \rightarrow \pi^- \nu$  the
$\pi^-$ is not isotropic in $\tau$ frame but rather travels in the backward
direction. (Note that polarization of the secondary resonances, $\rho^-$
and $a_1^-$, is not taken into account, {\it i.e.}, tertiary pions 
are assumed to be isotropic in the rest frame of the parent resonance.)
The lineshapes of intermediate resonances are generated in Gaussian forms.

     For a given decay channel of the $\tau$, the output of the generator
is a sequence of production--decay topologies (including tertiary pions from 
decays of secondary resonances) accompanied by weights that represent the 
computed matrix elements. Correspondingly, all distributions are plotted 
using these weights. At the moment, the individual hadrons of the primary 
hadron jet are not generated, and are treated as undetected. 

     Thus generated $\tau$ events are then propagated through the detector
using GEANT. Rather than fully digitize the response of electronic trackers
(drift chambers and silicon microstrips), the individual hits are randomly
distributed around central positions according to corresponding resolutions.
Then, particle momentum is obtained by fitting the "smeared" trajectory in
magnetic field. Likewise, the hits in individual cells of the EMCal are not 
fully simulated at this early stage. Instead, the energy and position of a 
shower are smeared according to aforementioned resolutions. The problem of 
overlapping showers is not yet addressed.

     For a given decay channel of the $\tau$, all histograms are normalized 
to the total number of {\it occurring} decays (so that the net contents are
just the accepted fraction of all $\tau$ decays in this channel). With this 
convention, the acceptance of a selection becomes immediately obvious without 
further normalization.
\subsection{Detecting the Decay  \xpidec}
     The momentum resolution for the $\pi^-$ with  $p_\pi > 1$ GeV from the 
decay \xpidec, as illustrated in Fig. \ref{momres}, on average amounts to 
some 2.7\% for either detector. The unsmeared transverse mass, 
$M_T = \sqrt{m_\pi^2 + p_T^2} + p_T$, is plotted
for all generated \xpidec\ events in Fig. \ref{m_t}. Also shown are the 
corresponding smeared distributions for those decay pions that pass the 
procedure-driven "default" selections in lab. momentum ($p_\pi > 1$ GeV), in 
emission angle relative to incident beam ($\Theta < 400$ mrad), in kink 
angle ($\theta_{\tau\pi} > 10$ mrad), and in transverse momentum to $\tau$ 
direction ($p_T > 250$ MeV). The Jacobian cusp of the original $M_T$
distribution is but slightly degraded by apparatus smearings of the
proposed COSMOS and TOSCA detectors. Therefore, in both detectors the
produced $\tau$ leptons will be efficiently tagged by \xpidec\ decays.
\subsection{Detecting the Decay \xrhodec, \ypi_gg}
     In this subsection that deals with the decay \xrhodec, the "default" 
selections for the $\pi^-$ from  \yrho_2pi\  (namely, $p_{\pi^-} > 1$ GeV,
$\Theta > 400$ mrad, $\theta_{\tau\pi} > 10$ mrad, and $p_T^\pi > 250$ MeV) 
are implicitly included in all distributions and quoted acceptances. For the 
$\pi^-$, we use the values of emission angles as measured near the production 
point in emulsion. 

     We assume that the conversion point of the photon from \ypi_gg\
is found and measured in emulsion and, therefore, its direction is precisely 
known. The energy of the parent photon is reconstructed by adding up the 
energies of unconverted daughter gammas (by hits in the EMCal) and of 
conversion electrons that are momentum-analyzed in the tracker.

     In this scheme of reconstruction, there is a potential danger of double
counting: a conversion electron, that has already been momentum-analyzed in 
the gap between the two targets, may shower in the next target producing 
further conversion electrons and gammas that should no longer be counted. We 
assume that, given a relatively large gap between successive targets, these 
"unwanted daughters" can be distinguished by their large displacement from 
the shower axis. Whether or not this can be done experimentally is currently 
under investigation, and our preliminary results are very encouraging. But 
for the moment, we realize this on GEANT level by "stopping" conversion 
electrons just before they hit the next target.

     By either allowing or excluding double counting, we obtain two different
estimates of the parent-photon energy. For either approach, the ratio
$E_\gamma^{vis} / E_\gamma^{true}$  between the estimated and true values of 
energy is plotted in Fig. \ref{ratio_1} for photons originating from different
emulsion targets of reconfigured COSMOS. Allowing "double counting" in 
$E_\gamma^{vis}$ is seen to result in overestimating the true energy for a 
significant fraction of photons from \ypi_gg. Again for either estimate of 
$E_\gamma^{vis}$, plotted in Fig. \ref{mgg_1} is the visible two-photon mass  
$m_{\gamma\gamma}$  for events occurring in different targets of COSMOS. Quite 
predictably, "double counting" in  $E_\gamma^{vis}$  degrades the $\pi^0$ 
signals in the two upstream targets. The very similar distributions of 
$E_\gamma^{vis} / E_\gamma^{true}$ and of $m_{\gamma\gamma}$ for different 
targets of reconfigured TOSCA are not illustrated. 

     With "double counting" excluded, the overall $m_{\gamma\gamma}$ 
distributions for the reconfigured COSMOS and TOSCA detectors are shown in 
Fig. \ref{mgg_2}. Fixing the mass window for $\pi^0$ selection requires 
detailed simulations of detector response to primary hadrons and to background 
processes, and therefore is impossible at this stage. For several realistic 
mass windows for $m_{\gamma\gamma}$ (135 $\pm$ 25, 20, and 15 MeV), the 
acceptances of reconfigured detectors to  \xrhodec\ decays are listed in 
Table \ref{rho_accept} together with those of the original COSMOS 
detector \cite{cosmos2}. (Note that reconstruction of \ypi_gg\ decays, that is 
essential for selecting the \xrhodec\ decays, is virtually impossible in the 
baseline design of TOSCA \cite{tosca}.) Either of the proposed detectors is
seen to have a substantially higher acceptance to  \xrhodec\ than the 
baseline detector of COSMOS.

     Tentatively selecting the \ypi_gg\ decays in the  $m_{\gamma\gamma}$ 
window of  $135 \pm 25$ MeV, we then plot the mass of the  $\pi^-\pi^0$ 
system, as reconstructed in either detector (see Fig. \ref{mgg_2}). And 
finally, in Fig. \ref{m_t} we plot the reconstructed transverse mass for 
\xrhodec, $M_T = \sqrt { m_{\pi\pi}^2 + p_T^2 } + p_T$. As the Jacobian 
structure of the original $M_T$ distribution is not destroyed by apparatus 
smearings, we may conclude that both proposed detectors will efficiently tag 
$\tau$ leptons by reconstructing the \xrhodec\ decays.
\subsection{Detecting the Decay \xa1dec, \ya1_3pi}
     This decay proceeds through an intermediate  $\rho^0$ 
($a_1^- \rightarrow \rho^0 \pi^-$) and has a branching fraction of 9.4\%.
The three-prong decay channel is plagued by the background from 
three-prong collisions of secondary hadrons with the nuclei of emulsion, 
including coherent production of the $a_1$ state by pions. Still, it may be 
worthwhile to check whether or not the expected $M_T$ distribution is 
distinctive. 

     Using the procedure-driven selections of $p_\pi > 1$ GeV and
$\Theta < 400$ mrad for all three pions and determining pion directions
near the production point in emulsion, in Fig. \ref{m_t} we plot the 
reconstructed transverse mass for the decay \xa1dec, \ya1_3pi\ :
$M_T = \sqrt { m_{3\pi}^2 + p_T^2 } + p_T$. (Note that no $\theta_{\tau\pi}$
selection is required for observing a three-prong decay in emulsion.) That 
the very distinct Jacobian structure of the original $M_T$ distribution 
persists in the smeared plots is very encouraging: backgrounds from pion 
dissociation and from coherent $a_1$ production should be amassed at much 
lower values of $M_T$. Whether or not the proposed detectors are good enough 
for separating the decays \xa1dec\ can only be decided by detailed 
comparisons with the simulated three-prong background.
\subsection{Detecting the decay $\tau^- \rightarrow e^- \nu \bar{\nu}$}
     Of the decays that are not in the (quasi-)two-body category, we 
consider the decay  \xedec\ that, compared to \xmudec, poses experimental
problems due to electrons showering in thick emulsion. Owing to thinner
emulsion targets and to good reconstruction of shower energy, we may detect
the \xedec\ decays more efficiently than the baseline apparata of COSMOS
and TOSCA.

     In reconstructing the electron from \xedec, we follow a strategy 
analogous to the one used for photons: the energy is reassembled from pieces,
using magnetic analysis for all positive and negative electrons leaving the 
emulsion and hits in the EMCal for the unconverted daughter gammas. Thus 
reconstructed energy is denoted as $E_e^{vis}$. Then, we are able to estimate 
the transverse momentum with respect to $\tau$ direction as 
$p_T^{vis} = E_e^{vis} \sin \theta_{\tau e}$, where $\theta_{\tau e}$
is the angle of the observed kink.

     That the sign of the generic electron be correctly assigned is important 
for rejecting the background from semileptonic decays of charm produced by
neutrinos. Therefore, the $e^-$ track must be followed in emulsion down to the 
exit and reliably linked to electronic trackers. For this to be possible, we 
assume that the energy of the generic-$e^-$ track at the exit point from 
the target of origin, $E_{track}^{exit}$, must exceed 1 GeV. In a thick 
target, the generic $e^-$ may fail this requirement by losing too much energy
through bremsstrahlung, even though the total energy of the shower can be
estimated.

     Thus, a candidate event for \xedec\ must satisfy the following conditions: 
$E_{track}^{exit} > 1$ GeV, $\Theta_e < 400$ mrad, $p_T^{vis} > 250$ MeV, 
and $\theta_{\tau e} > 10$ mrad. For the \xedec\ events that survive these
minimum selections in the detector, we plot the ratio between the 
reconstructed and original energies of the $e^-$, $E_e^{vis} / E_e^{true}$,
in Fig. \ref{ratio_2}. The reconstruction of electron energy is good in
both detectors proposed. For either the reconfigured and baseline designs of 
COSMOS and TOSCA, the fractions of all generated \xedec\ events that survive
the above selections are listed in Table \ref{e_accept}. The reconfigured 
detectors are seen to have higher acceptances to the decay \xedec\ than the 
baseline schemes for COSMOS and TOSCA.
\section{Summary}
     The large-$\Delta m^2$ domain of \omutau\ oscillations may be further
probed by the next generation of short-baseline neutrino experiments using 
the emulsion technique for detecting $\tau$ decays: COSMOS at Fermilab and
TOSCA at CERN. In this paper, we propose alternative 
schemes of these experiments that emphasize good spectrometry for charged 
particles and for electromagnetic showers and efficient reconstruction of 
\ypi_gg\ decays. Either configuration features a sequence of relatively thin 
emulsion targets, immersed in magnetic field and interspersed with
electronic trackers, and a fine-grained electromagnetic calorimeter built of 
lead glass. These elements act as an integral 
whole in reconstructing the electromagnetic showers. This conceptual scheme 
shows superior performance in identifying the $\tau$ (quasi-)two-body decays 
by their characteristic kinematics and in selecting the electronic decays of 
the $\tau$.
\newpage
\begin{table}
\begin{tabular}{|c|c|c|}
\hline
Quantity Compared     & COSMOS      & TOSCA \\
\hline
Proton energy, GeV    & 120               & 350 \\
Protons per cycle     & $4\cdot10^{13}$   & $6\cdot10^{13}$ \\
Cycle time, s         & 1.9                & 19.2 \\
Protons per year      & $3.7\cdot10^{20}$  & $4.3\cdot10^{19}$ \\
Length of decay channel, m                & 800       & 414 \\
Distance from target to detector, m       & 960       & 806 \\
\hline
Emulsion mass, kg                  & 865      & 2784 \\
Emulsion area, m$^2$               & $1.80\times1.40$  & $1.44\times1.44$ \\
Emulsion thickness, cm             & $2\times4.5$      & $6\times6.0$ \\
\hline
$\langle E \rangle$ of $\nu_\mu$ beam (by flux), GeV   & 12        & 25 \\
$\langle E \rangle$ of $\nu_\mu$-induced CC events, GeV  & 18       &37 \\
$\langle E \rangle$ of $\nu_\tau$-induced CC events, GeV & 24       &54 \\
$\langle L / E \rangle$, km/GeV                      & 0.023      & 0.012 \\
$\langle\sigma(\nu_\tau)\rangle / \langle\sigma(\nu_\mu)\rangle$ 
                                                     & 0.27       & 0.50 \\
\hline
$\Sigma$ (BR $\times$ efficiency)          & 0.114         & 0.104 \\
Total number of $\nu_\mu$-induced CC events 
                                         & $8.1\cdot10^6$  & $6.0\cdot10^6$ \\
Total number of background events        & 1.3            & 1.3 \\
Total background events, 90\% CL         & 3.5            & 3.5 \\
Reach in $\sin^2(2\theta)$ for large $\Delta m^2$ 
                                   & $2.8\cdot10^{-5}$  & $1.5\cdot10^{-5}$ \\
Reach in $\Delta m^2$ for maximum mixing, eV$^2$   & 0.10    & 0.10 \\
\hline
\end{tabular}
\caption{General parameters of the baseline designs of COSMOS \cite{cosmos2} 
and TOSCA \cite{tosca}.}
\label{compare}
\end{table}
\newpage
\begin{table}
\begin{tabular}{|c|c|c|c|}
\hline
Mass window for         & Our design    & Our design    & Baseline design \\
selecting the $\pi^0$   & for COSMOS    & for TOSCA     & for COSMOS \\
\hline
Unrestricted           & 0.40                  & 0.42                 & 0.31 \\
$135 \pm 25$ MeV       & 0.31                  & 0.28                 & 0.19 \\
$135 \pm 20$ MeV       & 0.29                  & 0.25                 & 0.18 \\
$135 \pm 15$ MeV       & 0.26                  & 0.21                 & 0.16 \\
\hline
\end{tabular}
\caption{The fractions of all \xrhodec, \yrho_2pi\ decays that survive 
different $\pi^0$ selections (as indicated) and the "default" selections
for the $\pi^-$.}
\label{rho_accept}
\end{table}     
\clearpage
\begin{table}
\begin{tabular}{|c|c|c|c|}
\hline
  Our design    & Our design      & Baseline design   & Baseline design \\
  for COSMOS    & for TOSCA       & for COSMOS        & for TOSCA \\
\hline
      0.55            & 0.56               & 0.31              & 0.42 \\
\hline
\end{tabular}
\caption{The fractions of all \xedec\ decays that survive the minimum
selections in the detector, as detailed in the text.}
\label{e_accept}
\end{table}     
\clearpage
{\bf Figure Captions}

\begin{figure}
\vspace{19 cm}
\includegraphics{pict2cosmos.ps} 
\includegraphics{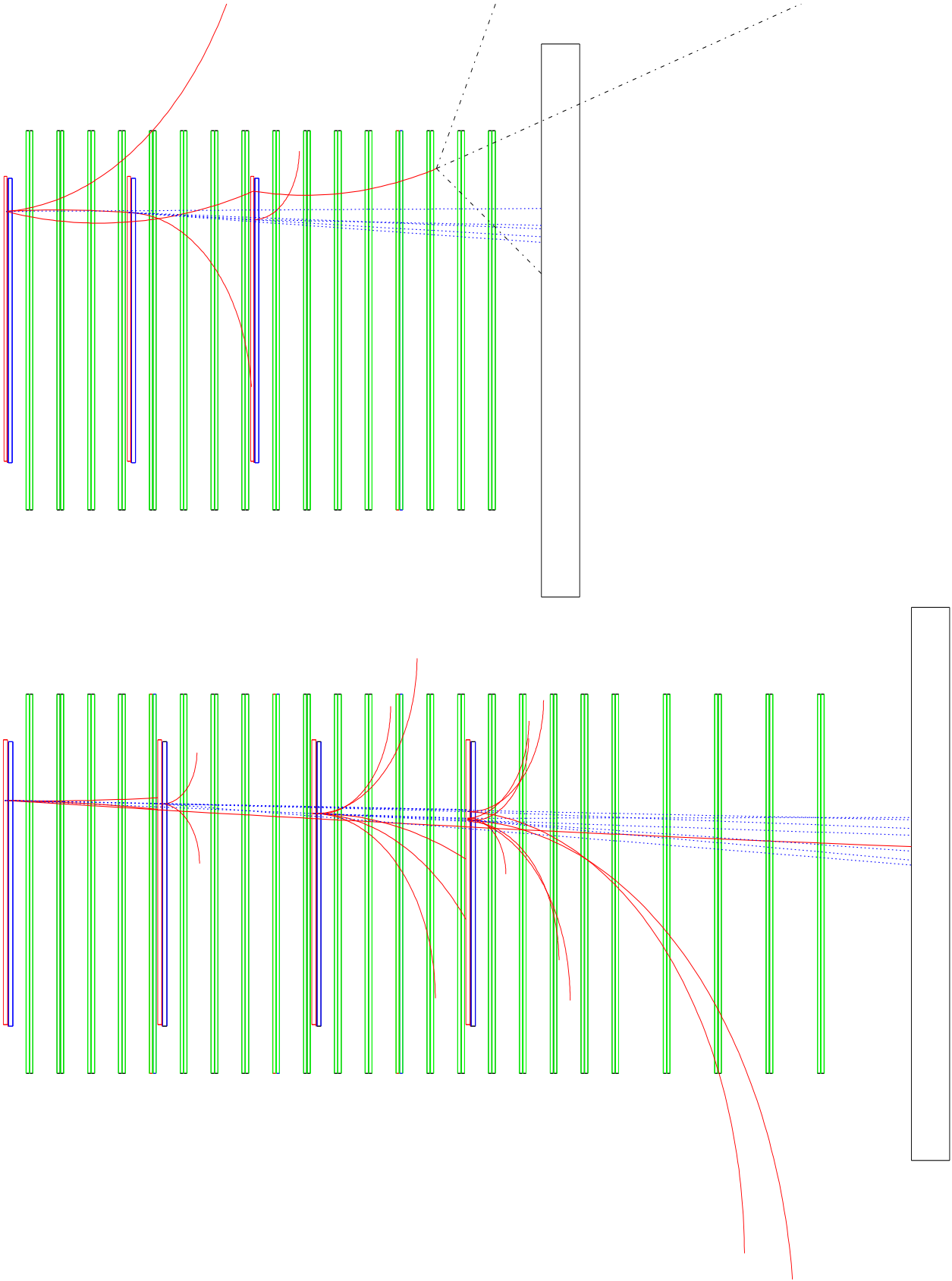} 
\caption{The configurations proposed for COSMOS {\bf(top)} and TOSCA 
{\bf(bottom)}. Overlaid are the \xrhodec\ decays in either detector.}
\label{detectors}
\end{figure}

Fig. \ref{detectors}. The configurations proposed for COSMOS {\bf(top)} and 
TOSCA {\bf(bottom)}. Overlaid are the \xrhodec\ decays in either detector.

\begin{figure}
\vspace{19 cm}
\includegraphics{enua.ps} 
\includegraphics{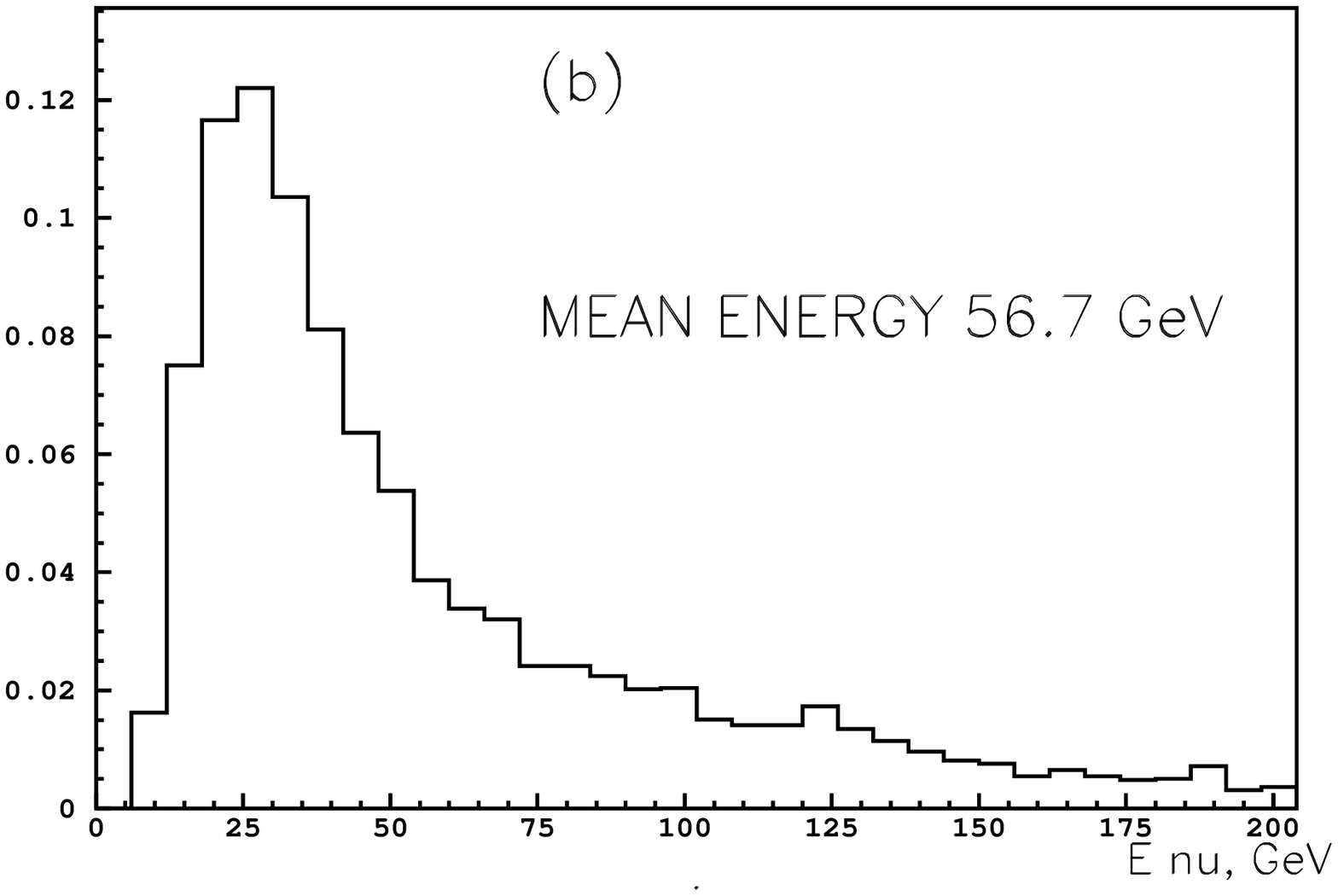} 
\caption{The computed energy spectra of  $\nu_\tau$-induced CC events in
COSMOS {\bf(a)} and TOSCA {\bf(b)}. The absolute normalization is arbitrary.}
\label{cc_spectra}
\end{figure}

Fig. \ref{cc_spectra}. The computed energy spectra of  $\nu_\tau$-induced 
CC events in COSMOS {\bf(a)} and TOSCA {\bf(b)}. The absolute normalization 
is arbitrary.

\begin{figure}
\vspace{19 cm}
\includegraphics{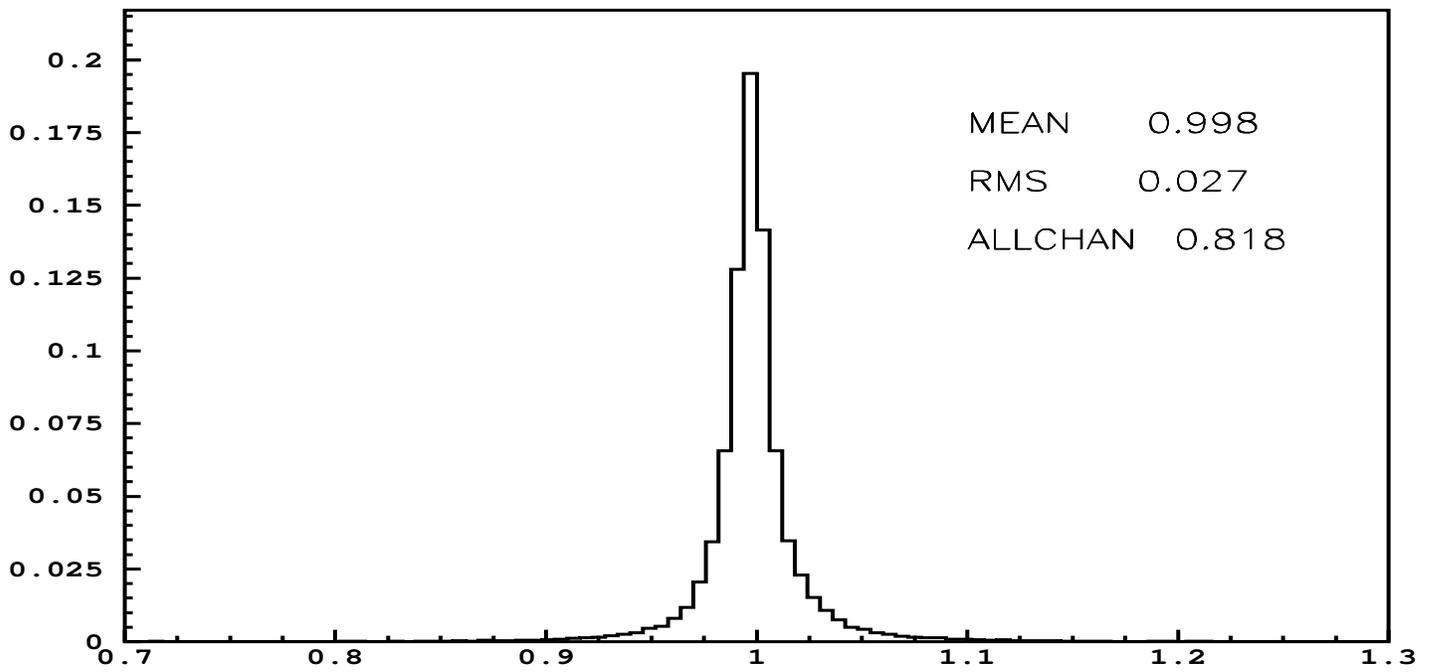} 
\includegraphics{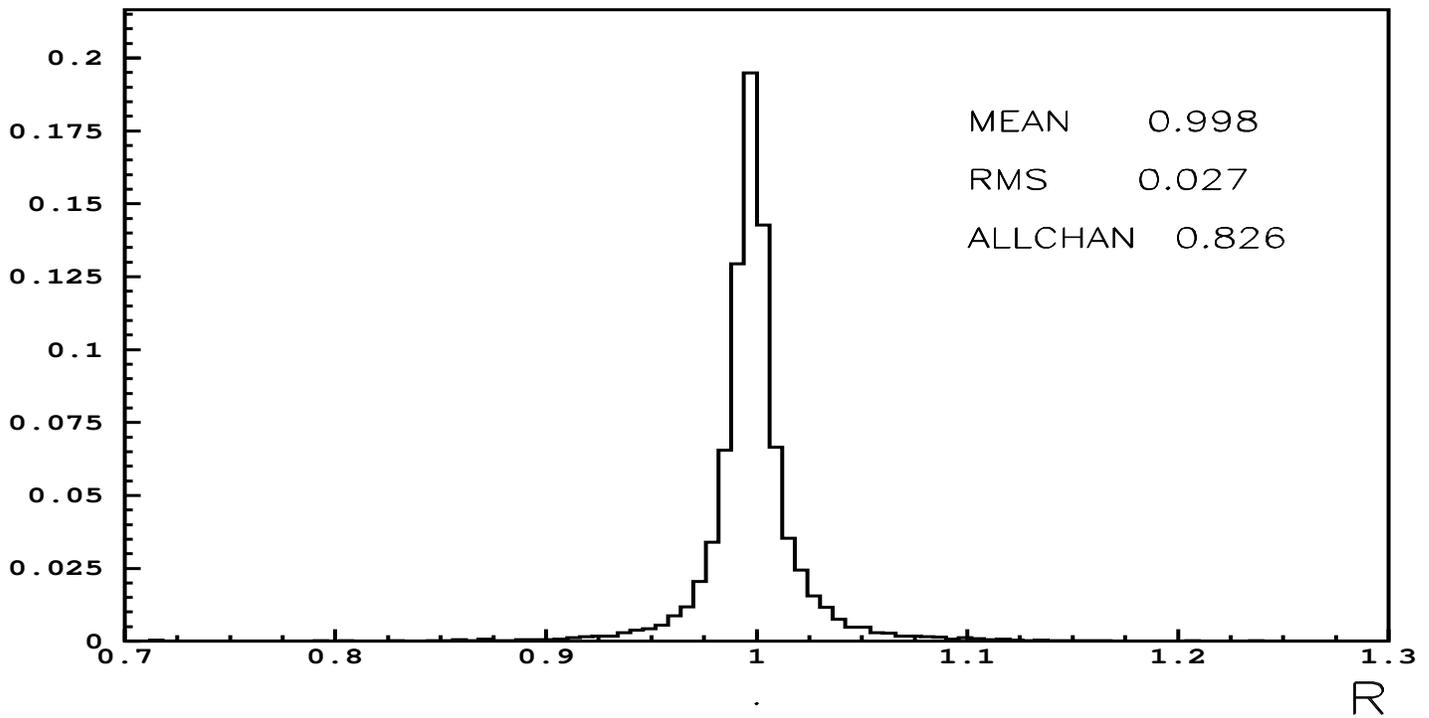} 
\caption{In the decay \xpidec, the ratio between the measured and true 
values of $\pi^-$ momentum, $R=p_\pi^{vis} / p_\pi^{true}$, for the proposed
schemes of COSMOS {\bf(a)} and TOSCA {\bf(b)}. Here and in all subsequent
figures, the smearing due to Coulomb scattering in emulsion is fully taken 
into account.}
\label{momres}
\end{figure}

Fig. \ref{momres}. In the decay \xpidec, the ratio between the measured and 
true values of $\pi^-$ momentum, $R=p_\pi^{vis} / p_\pi^{true}$, for the 
proposed schemes of COSMOS {\bf(a)} and TOSCA {\bf(b)}. Here and in all 
subsequent figures, the smearing due to Coulomb scattering in emulsion is 
fully taken into account.

\begin{figure}
\vspace{19 cm}
\includegraphics{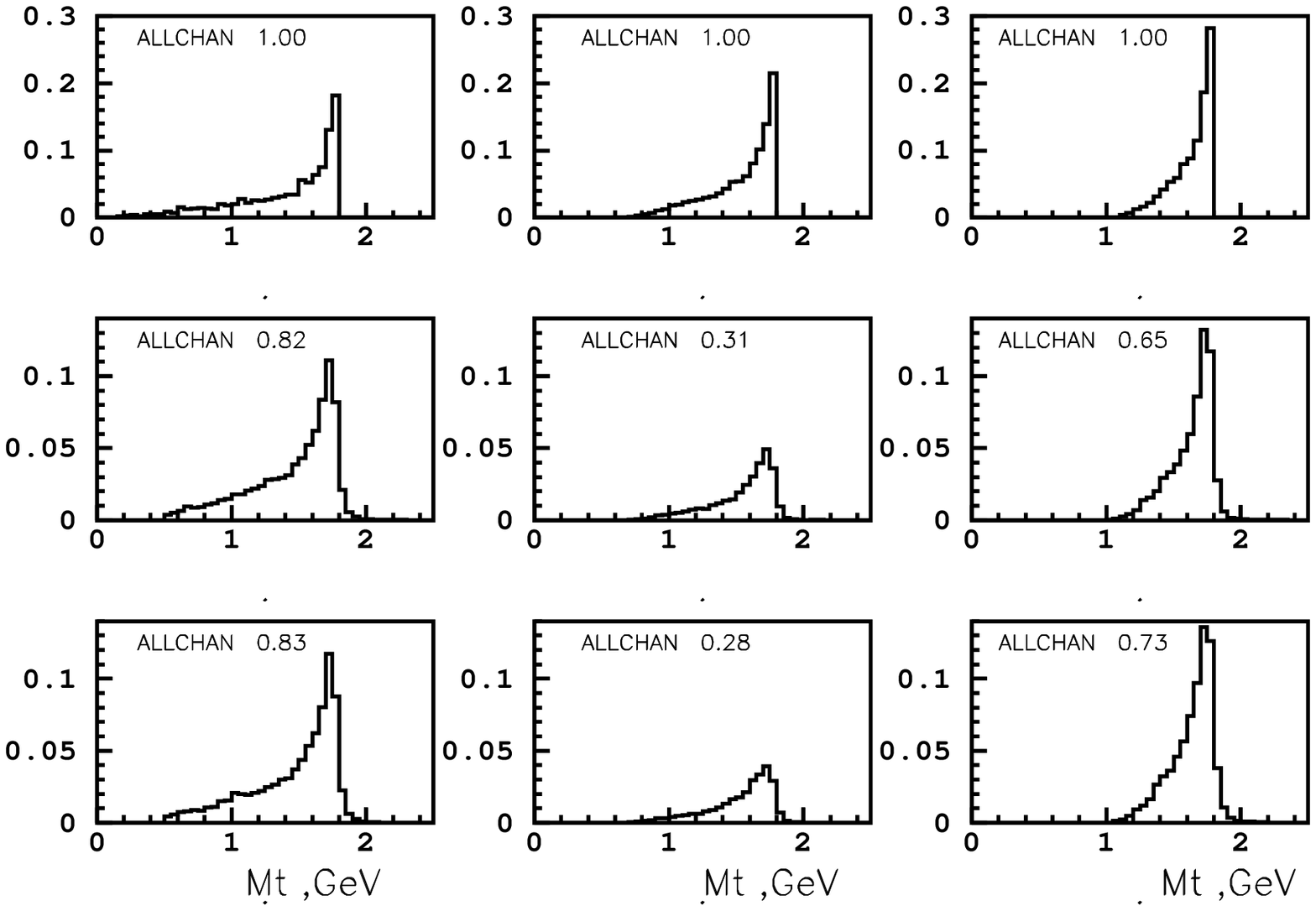} 
\caption{Transverse mass  $M_T = \sqrt{p_T^2 + m_h^2} + p_T$  for the
(quasi-)two-body decays  $\tau^- \rightarrow h^-\nu$  with $h^- = \pi^-$ 
(left-hand column), $h^- = \rho^-\rightarrow\pi^-\pi^0$ (middle column), and
$h^- = a_1^-\rightarrow\pi^-\pi^+\pi^-$ (right-hand column). The unsmeared
$M_T$ distributions for all generated events in each channel are shown in the 
top row. The smeared distributions for events detected by the reconfigured
COSMOS and TOSCA are shown in the middle and bottom rows, respectively.}
\label{m_t}
\end{figure}

Fig. \ref{m_t}. Transverse mass  $M_T = \sqrt{p_T^2 + m_h^2} + p_T$  for the
(quasi-)two-body decays  $\tau^- \rightarrow h^-\nu$  with $h^- = \pi^-$ 
(left-hand column), $h^- = \rho^-\rightarrow\pi^-\pi^0$ (middle column), and
$h^- = a_1^-\rightarrow\pi^-\pi^+\pi^-$ (right-hand column). The unsmeared
$M_T$ distributions for all generated events in each channel are shown in the 
top row. The smeared distributions for events detected by the reconfigured
COSMOS and TOSCA are shown in the middle and bottom rows, respectively.

\begin{figure}
\vspace{19 cm}
\includegraphics{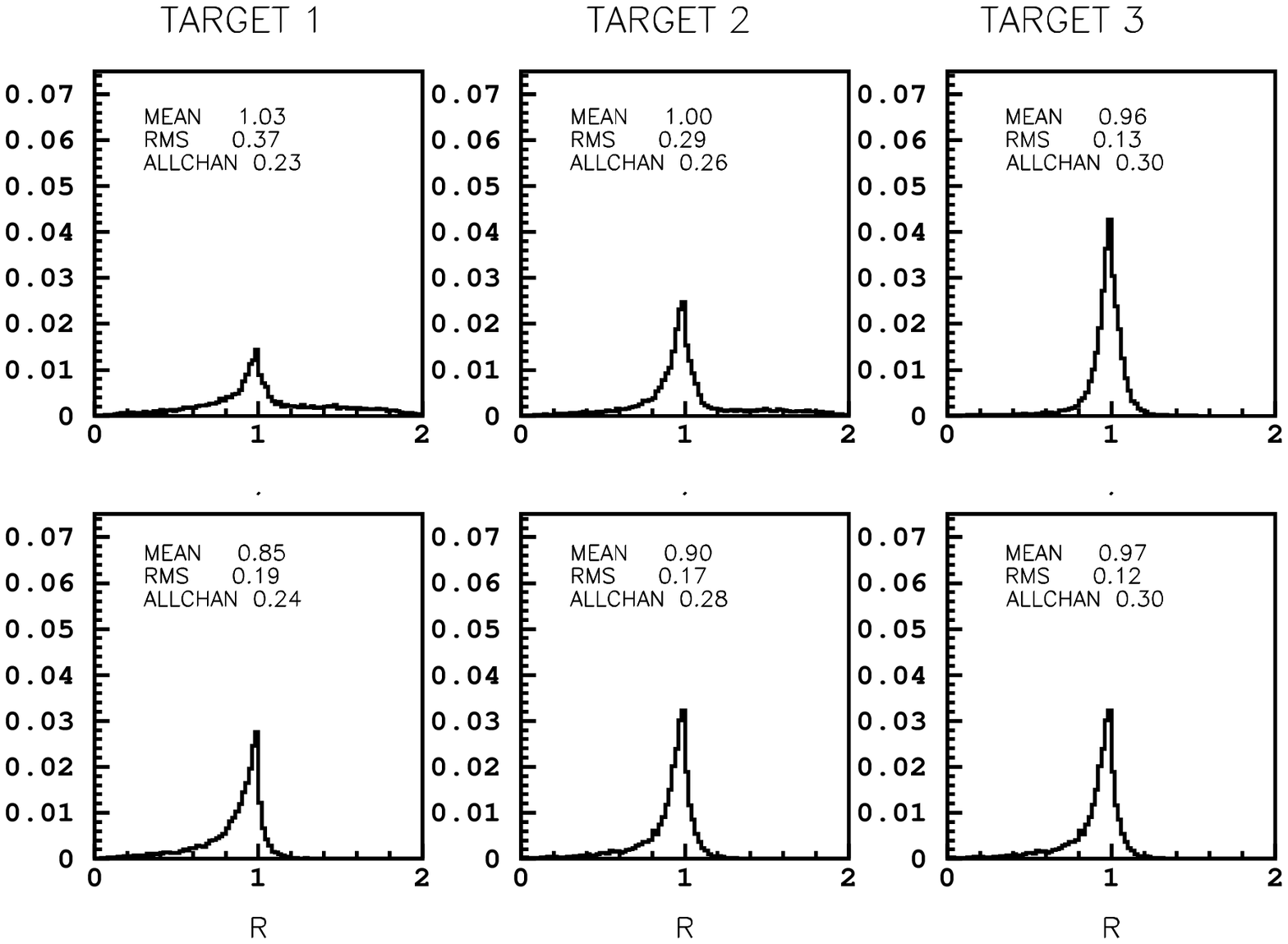} 
\caption{The ratio $R=E_\gamma^{vis} / E_\gamma^{true}$  between the estimated 
and true values of energy for photons originating from different emulsion 
targets of reconfigured COSMOS, with "double counting" in $E_\gamma^{vis}$ 
either allowed (the top row) or forbidden (the bottom row). The three columns 
stand for the three targets of the proposed detector.}
\label{ratio_1}
\end{figure}

Fig. \ref{ratio_1}. The ratio $R=E_\gamma^{vis} / E_\gamma^{true}$ 
 between the 
estimated and true values of energy for photons originating from different 
emulsion targets of reconfigured COSMOS, with "double counting" in 
$E_\gamma^{vis}$ either allowed (the top row) or forbidden (the bottom row). 
The three columns stand for the three targets of the proposed detector.

\begin{figure}
\vspace{19 cm}
\includegraphics{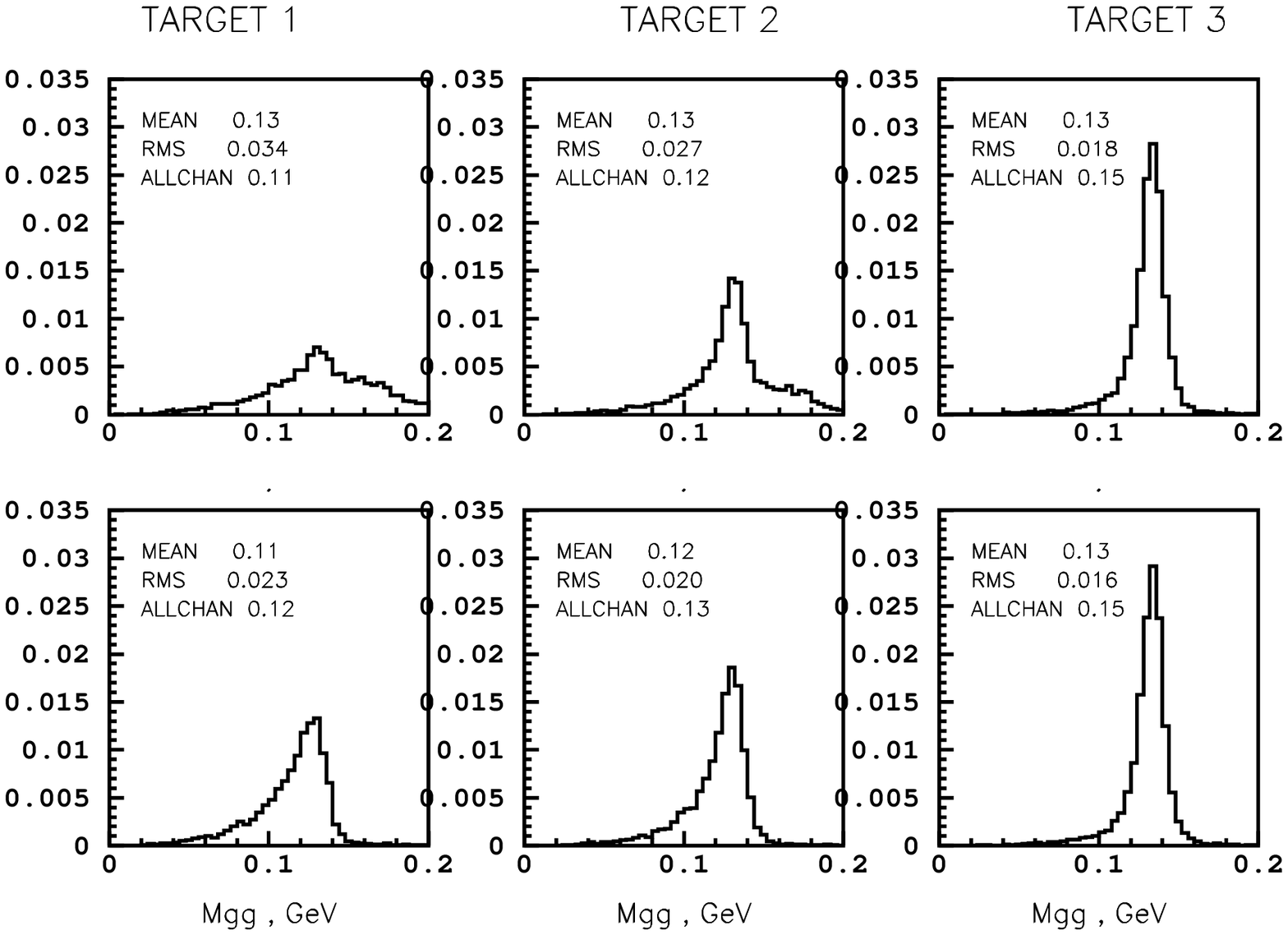} 
\caption{The reconstructed two-photon mass  $m_{\gamma\gamma}$ for the \ypi_gg\
decays occurring in different emulsion targets of reconfigured COSMOS, with 
"double counting" in $E_\gamma^{vis}$ either allowed (the top row) or forbidden (the 
bottom row). The three columns stand for the three targets of the proposed 
detector.}
\label{mgg_1}
\end{figure}

Fig. \ref{mgg_1}. The reconstructed two-photon mass  $m_{\gamma\gamma}$ for 
the \ypi_gg\ decays occurring in different emulsion targets of reconfigured 
COSMOS, with "double counting" in $E_\gamma^{vis}$ either allowed (the top 
row) or forbidden (the bottom row). The three columns stand for the three 
targets of the proposed detector.

\begin{figure}
\vspace{19 cm}
\includegraphics{pmasa.ps} 
\includegraphics{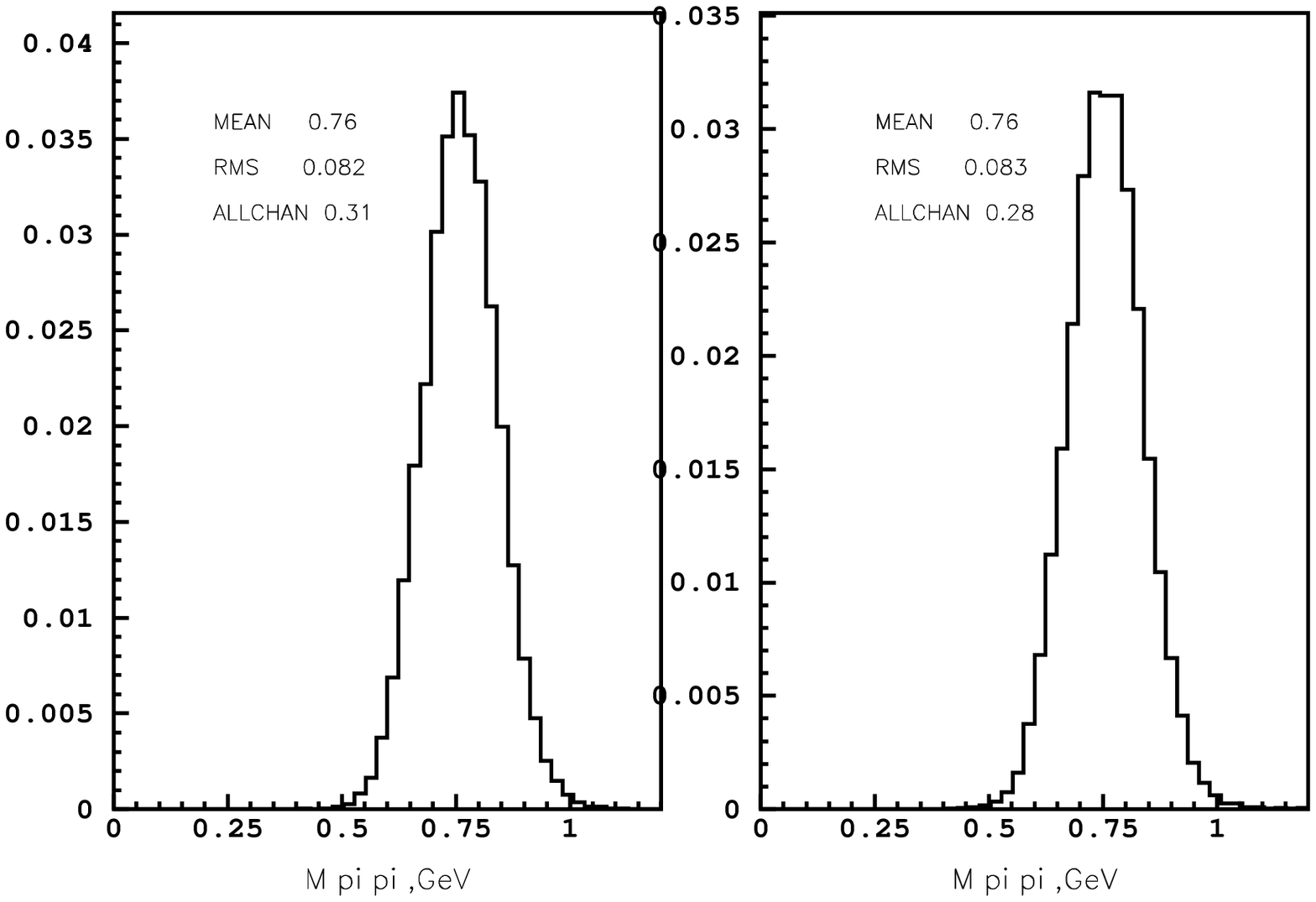} 
\caption{The masses $m_{\gamma\gamma}$ (top row) and  $m_{\pi\pi}$
(bottom row), as reconstructed in the reconfigured COSMOS  and TOSCA detectors
(the left-hand and right-hand columns, respectively) excluding double counting 
in $E_\gamma^{vis}$. Note that the lineshape of the original  $\rho^-$  has 
been generated as a Gaussian with  $\sigma = 75$ MeV.}
\label{mgg_2}
\end{figure}

Fig. \ref{mgg_2}. The masses $m_{\gamma\gamma}$ (top row) and  $m_{\pi\pi}$
(bottom row), as reconstructed in the reconfigured COSMOS  and TOSCA detectors
(the left-hand and right-hand columns, respectively) excluding double counting 
in $E_\gamma^{vis}$. Note that the lineshape of the original  $\rho^-$  has 
been generated as a Gaussian with  $\sigma = 75$ MeV.

\begin{figure}
\vspace{19 cm}
\includegraphics{evisa.ps} 
\includegraphics{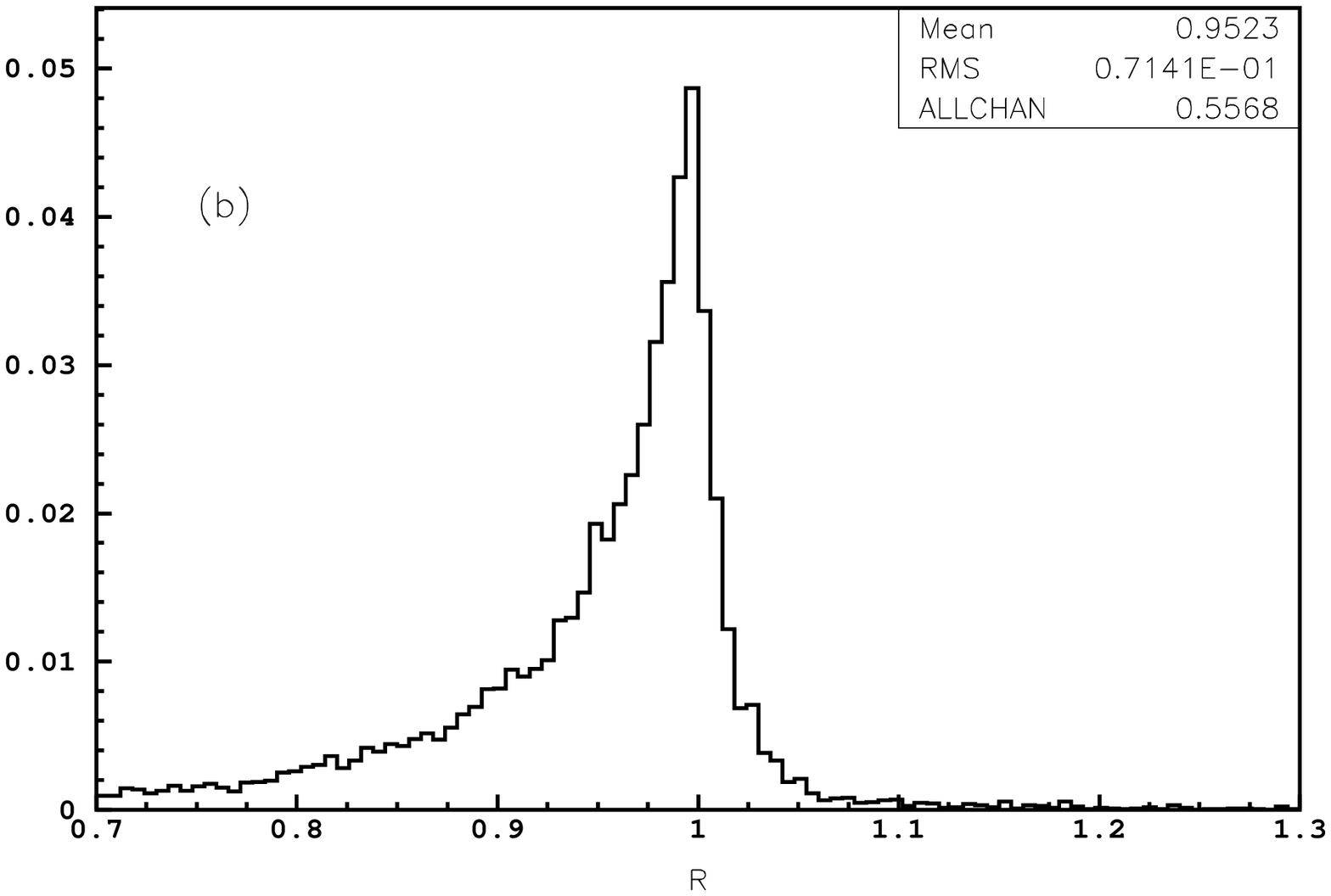} 
\caption{The ratio between the reconstructed and original energies of the 
$e^-$ from \xedec, $R=E_e^{vis} / E_e^{true}$, for the proposed configurations 
of COSMOS {\bf(a)} and TOSCA {\bf(b)}.}
\label{ratio_2}
\end{figure}

Fig. \ref{ratio_2}. The ratio between the reconstructed and original energies 
of the $e^-$ from \xedec, $R=E_e^{vis} / E_e^{true}$, for the proposed 
configurations of COSMOS {\bf(a)} and TOSCA {\bf(b)}.
\newpage
\newpage
\newpage
\newpage
\newpage
\newpage
\newpage
\newpage
\end{document}